\documentclass[preprints,article,accept,moreauthors]{mdpi}

\firstpage{1} 
\pubvolume{xx}
\issuenum{1}
\articlenumber{5}
\pubyear{2019}
\copyrightyear{2019}
\history{Received:  29 November 2019; Accepted: 7 January 2020; Published: 10 January 2020}

\pdfoutput=1

\usepackage{amsmath}
\usepackage{amsfonts}
\usepackage{amssymb}

\newcommand{\ce}{\varepsilon}
\newcommand{\pd}{\partial}
\newcommand{\ad}{\operatorname{ad}}
\Title{GRAVITY AND NONLINEAR SYMMETRY REALISATION}

\Author{Andrej Arbuzov $^{1,3,\dagger}$\orcidA{}, Boris Latosh $^{2,3\ddagger}$\orcidB{}}

\AuthorNames{Andrej Arbuzov, Boris Latosh}

\address{%
$^{1}$ \quad Bogoliubov Laboratory for Theoretical Physics, JINR, Dubna 141980, Russia;\\
$^{2}$ \quad Department of Physics and Astronomy, University of Sussex, Brighton, BN1 9QH, United Kingdom;\\
$^{3}$ \quad Dubna State University, Universitetskaya str. 19, Dubna 141982, Russia}

\firstnote{arbuzov@theor.jinr.ru} 
\secondnote{b.latosh@sussex.ac.uk}

\abstract{Application of nonlinear symmetry realisation technique to gravity is studied. We identify the simplest extensions of the Poincare group suitable for nonlinear realisation at the level of physical fields. Two simple models are proposed. The first one introduces additional scalar degrees of freedom that may be suitable for driving inflation. The second one describes states with well-defined mass that lack a linear interaction with matter states. We argue that this phenomenon points on a necessity to draw a distinction between gravitational states with well defined masses and states that participate in interaction with matter.}



\begin{document}

\section{Introduction}\label{Introduction}

General Relativity (GR) is widely accepted as the best theory of gravity. At the scale of star systems GR provides the excellent fit for observational data~\cite{Will:2014kxa,Berti:2015itd,Ashby:2003vja}. At the cosmological scale GR consistently describes a wast array of phenomena~\cite{Gorbunov:2011zz,Gorbunov:2011zzc,WeinbergSteven2008C}, however, phenomena of dark matter and the late-time accelerated expansion (dark energy) challenge GR applicability at this scale \cite{Hinshaw:2012aka,Ade:2015xua,Clowe:2006eq,Moore:1999nt,Davis:1985rj}. Although they can be described within GR, their true physical nature remains unknown. A consistent cosmological model also requires an inflationary phase of expansion which lies beyond GR's applicability~\cite{Linde:2014nna,Starobinsky:1980te,Nojiri:2017ncd}. These facts provide a ground to consider GR as a effective gravity model relevant for small spacial and energy scales. Therefore, it is required to study alternative gravity models that modify GR and may be relevant for all spacial and energy scales.

Classical (non-quantum) gravity models can be classified via the Lovelock theorem~\cite{Lovelock:1972vz,Lovelock:1971yv}. The theorem states that the left hand side of the Einstein equations is defined uniquely, if certain conditions are met. These conditions are related to the number of spacetime dimensions, the differential order of the field equations, and the use of a symmetric rank-2 tensor as a dynamic variable. To obtain an alternative gravity model one should violate any of these conditions. Therefore, classical gravity models can be classified in accordance with the Lovelock theorem conditions which they violate~\cite{Berti:2015itd,Sotiriou:2015lxa}. For instance, $f(R)$ gravity models violate the theorem conditions as they have higher order field equations; scalar-tensor gravity models violate the conditions as they use an additional scalar field that is not related with the metric tensor.

The main disadvantage of such a classification is the fact that it deals with a mathematical way to construct a Lagrangian, but does not account for its physical features. At the same time models which violate different conditions of the Lovelock theorem may share some physical features. The best example is given by $f(R)$ gravity and scalar-tensor models, as these models introduce one additional scalar gravitational degree of freedom. To be exact, $f(R)$ gravity models can be mapped onto a particular class of scalar-tensor models~\cite{Rodrigues:2011zi,DeFelice:2010aj,Sotiriou:2008rp}. Therefore, one can classify gravity models with respect to their perturbation spectra. 

Study of gravity models based on their perturbation spectra goes in line with the recent development of the effective field theory (EFT) for gravity~\cite{Burgess:2003jk,Donoghue:1994dn,Donoghue:2012zc,Calmet:2017voc,Magnano:1995pv}. The EFT approach is aimed at creation of a quantum gravity model relevant for the low-energy regime. Within EFT small gravitational field perturbations are quantised and treated as a low-energy manifestation of the true quantum gravitational field. In such a way an EFT gravity model requires data on the perturbation spectrum of the correspondent classical gravity model. In this context classification of gravity models via their perturbation spectra is more suitable, as it complements the EFT approach. 

The other major advantage of such an approach is the fact that it provides a way to study symmetry of the gravity sector.
There are two ways to introduce additional symmetry to a gravity model. First one is to introduce an additional symmetry at the coordinate level; second one is to introduce an additional symmetry at the level of physical fields, such as metric (or vierbein), vectors, fermions, and scalars. These opportunities are heavily constrained by multiple factors which we discuss further. However, the later opportunity is less constrained and proven to be more fruitful in gravity study.
A simple example of such a model is given by the so-called (generalised) Galileons~\cite{Deffayet:2013lga,Kobayashi:2019hrl}. Originally Galileon models describe a scalar field with the so-called generalised Galilean symmetry in a flat spacetime. Later these models were generalised for a curved spacetime and it was proven that they form a class of the most general scalar-tensor gravity models with second order field equations (they are also known as Horndeski models)~\cite{Kobayashi:2011nu,Horndeski:1974wa}. Therefore, (generalised) Galileon models prove that introduction of a new symmetry can significantly affect a gravity model.

Opportunity to introduce new symmetries in the gravity sector are heavily constraint. At the level of coordinates only infinite-dimensional representations of Lorentz group can take place \cite{Weinberg:1995mt}. Otherwise, a correspondent quantum field theory lacks unitarity and cannot be considered relevant. At the level of physical fields Lorentz group can have a finite-dimensional representation, as it does not affect an underlying quantum theory (as it is pointed out in \cite{Weinberg:1995mt}). However, physical fields must be subjected to a realisation of Poincare algebra. Poincare algebra allows one to define mass and spin operators consistently \cite{Wigner:1939cj}, as they are related with correspondent Casimir operators.

If a new symmetry is introduced at the coordinate level, it may not be possible to construct a suitable infinite-dimensional Lorentz group representation and reproduce required tools of Riemannian geometry. If a new symmetry is introduced at the level of fields, it may not be possible to define mass and spin operators. This, in turn, makes it impossible to define states with definite mass and spin and ruins a possibility to justify the standard quantum field theory formalism. An explicit example of such a phenomenon is given by the conformal symmetry, as the conformal group has a different number of Casimir operators~\cite{Lagu:1974aj}. Therefor, in order to preserve the conventional quantum field theory one must preserve a linear realisation of the Poincare group in the gravitational sector.

The only possible way to extend symmetry of the gravity sector is to introduce a new symmetry realised in a nonlinear way. The physical reasons behind that assumption are related to the EFT approach. As we mentioned before, EFT deals with a low energy manifestation of gravity. It is known that some physical systems can experience a spontaneous symmetry breaking in the low energy regime. Namely, the Higgs sector of the standard model of particle physics provides the best known example. The Higgs field admits the $SU(2)\times SU(2)$ symmetry, but in the low energy regime it develops a ground state that spontaneously breaks this symmetry~\cite{Higgs:1964pj,Weinberg:1967tq,Greiner:1993qp}.

In full analogy with the Higgs field, a gravity field may experience spontaneous breaking of a bigger symmetry down to the Poincare group in the low energy regime. Because of this all physical states respect only the Poincare symmetry alongside with the correspondent Casimir operators. Various string inspired models support that idea, for instance, AdS/CFT correspondence provides an explicit way to relate a conformal particle model to a higher dimensional gravity model~\cite{Witten:1998qj,Maldacena:1997re,Gubser:1999vj}.

In this paper we discuss one particular way to introduce a new symmetry at the level of physical fields in the gravity sector via the nonlinear symmetry realisation technique. This technique was developed in papers~\cite{Coleman:1969sm,Callan:1969sn,Weinberg:1978kz} and widely applied for various physical problems. 
In particular, the following fundamental results were obtained via this technique. First, Ogievetsky theorem~\cite{Ogievetsky:1973ik} established that infinite-dimensional algebra of coordinate transformation can be obtained as a closure of two finite-dimensional algebras. Second, it was shown, that nonlinear realisation technique allows one to restore the whole framework of Riemannian geometry ~\cite{Borisov:1974bn}.

In other words, results \cite{Ogievetsky:1973ik,Borisov:1974bn} show, that a nonlinear realisation of particular symmetries already takes place at the level of coordinates. This gives one a ground to expect a nonlinear realisation of the same symmetries at the level of physical fields. The aim of our paper is to investigate, if an implementation of nonlinear symmetry realisation at the level of physical fields may produce any modification of GR suitable for further study.
For our purposes we use a technique that realises a symmetry group on a quotient space. Then we use the Ogievetsky theorem~\cite{Ogievetsky:1973ik} to define the simplest modifications of the Poincare group which can be used with the nonlinear symmetry realisation technique. 
We discuss possible implications of the nonlinear symmetry realisation for gravity models and find direct physical corollaries of these implications. This study allows us to highlight the most perspective directions of further research and possible consequences of an implication of nonlinear symmetry realisation technique.

This paper is organised as follows. In Section~\ref{many_faces_of_nonlinear_symmetry_realisation} we discuss relations between nonlinear symmetry realisation and spontaneous symmetry breaking. This discussion is due, as it allows us to justify applications of nonlinear symmetry realisation and the usage of the Goldstone theorem. In Section~\ref{Nonlinear_symmetry_realisation_quotient} we briefly present the mathematical theory of nonlinear group realisation on a quotient. In Section~\ref{Implementation} we implement this technique and discuss its physical corollaries. 
We also compare our approach to the classical results \cite{Borisov:1974bn} and discuss their relations. In Section~\ref{Discussion} we summarise our results and present conclusions on the possibility to implement the nonlinear symmetry realisation technique. Appendix~\ref{appendix_algebras} contains data on the Lie algebras that were used in the paper.

\section{Many faces of nonlinear symmetry realisation}\label{many_faces_of_nonlinear_symmetry_realisation}

Nonlinear symmetry realisation and spontaneous symmetry breaking refer to the same class of physical phenomena may be except for a few special cases, like spontaneous symmetry breaking of linear groups. When these notions refer to the same class of phenomena, they simply highlight different features of the same phenomena. 

Traditionally, it is said that a symmetry group $G$ of a physical system is spontaneously broken down to a group $H$ if the system enters a state that is invariant with respect to $H$, but not invariant with respect to $G$~\cite{Weinberg:1996kr,Nair:2005iw,ItzyksonC1980QFT}. A simple example of such a phenomenon is given by the Higgs sector of the standard model. The potential of the Higgs field is invariant with respect to $SU(2)\times SU(2)$ group, but any particular ground state is not.

It is crucial to highlight that the bigger symmetry group $G$ is a property of a system as a whole and it cannot be changed or affected by matter. For instance, the potential of the Higgs field always respect the global $SU(2)\times SU(2)$ symmetry as this symmetry is a fundamental property of the Higgs field itself~\cite{Elitzur:1975im,Chernodub:2008rz}. On the other hand, the smaller symmetry group $H$ is a property of a particular matter configuration, so it cannot be considered as fundamental. In accord with this logic the smaller (custodial) symmetry of the Higgs sector manifests itself at the level of perturbation phenomena. The perturbation theory of the Higgs field is based around a ground state which is not invariant with respect to $SU(2)\times SU(2)$. In such a way the global symmetry is broken due to the existence of a particular matter configuration around which the perturbation theory is constructed. Therefore, when a symmetry $G$ is broken down to $H$, the bigger symmetry $G$ remains a fundamental property of the system. But a manifestation of the bigger symmetry is nonlinear, as matter must respect the smaller symmetry group $H$. 

As it was noted, there are a few special cases, when spontaneous symmetry breaking lies beyond nonlinear symmetry realisations. There are models with spontaneously breaking of Abelian group of constant field shifts, which cannot be treated as nonlinear symmetry realisation (see \cite{Narnhofer:1999iv} and references therein). This is due to the fact, that one requires a non-trivial subgroup to construct a nonlinear symmetry realisation. In case of linear (one-dimensional) Abelian groups there are none such groups, so one cannot apply nonlinear symmetry realisation technique. Nonetheless, this case present one particular exception irrelevant for the study presented in this paper. Therefore we do not discuss it further.

This proves the original claim that spontaneous symmetry breaking and nonlinear symmetry realisation are different names of the same phenomenon, except for a few special cases. The notion of spontaneous symmetry breaking is used when it is required to highlight the fact that the matter respects only a smaller symmetry group. The notion of nonlinear symmetry realisation is used when it is required to highlight the fact that the original bigger symmetry group still manifests itself.

Such a relation between these notions can be illustrated with the so-called inverse Higgs effect \cite{Ivanov:1975zq} (see also \cite{Ivanov:2016lha} for a more detailed review). This effect takes place when a $D$ brane is embedded in a multidimensional space. The multidimensional Lorentz group is broken down due to the existence of the brane. But at the same time the multidimensional Lorentz symmetry manifests itself on the brane in a nonlinear way.

The simplest case of the inverse Higgs effect corresponds to a one dimensional world line of a single massive particle in a two dimensional spacetime. The existence of a massive particle breaks the boost component of the correspondent Lorentz group. It is shown in~\cite{Ivanov:1999gy} that implications of nonlinear symmetry realisation methods allow one to restore the Lagrangian of a free massive particle. At the same time the same physical system can be treated as a system with spontaneously broken symmetry. The system admits a continuous manifold of free massive particle world lines and each world line is invariant under a certain subgroup of the two dimensional spacetime Lorentz group. Therefore the correspondent Lorentz group is spontaneously broken.

The fact that nonlinear symmetry realisations and spontaneous symmetry breaking refer to the same phenomenon allows one to use the Goldstone theorem. Namely, if a bigger symmetry group $G$ is realised on a smaller group $H$ and a quotient $G/H$, then the correspondent model has $\dim(H)$ massless 
degrees of freedom~\cite{Nair:2005iw}. This allows one to conjecture that gravitational degrees of freedom are massless due to the Goldstone theorem. In other words, we conjecture that gravitons are goldstone bosons and the gravity itself appears due to a nonlinear manifestation of some symmetry. The same conjecture was used to justify certain gravity models earlier~\cite{Borisov:1974bn}.

\section{Nonlinear symmetry realisation via group action on a quotient} \label{Nonlinear_symmetry_realisation_quotient}

From the mathematical point of view any $n$-dimensional real Lie group $G$ is a real continuous $n$-dimensional manifold. Therefore any element $g$ of such a group $G$ can be parameterised by $n$ real numbers $\zeta_k$:
\begin{align}
  g=g(\zeta_1,\cdots,\zeta_n).
\end{align}
Numbers $\zeta_k$ form coordinates on the group manifold $G$. Generators of the Lie group $G$ should be treated as vectors tangent to the group $G$ at the point corresponding to the identical transformations. It is always possible to set the frame origin at the point of the identical transformation and define the group generators as follows:
\begin{align}
  I_k \overset{\text{def}}{=} \left. \cfrac{\pd}{\pd \zeta_k} g(\zeta_1,\cdots,\zeta_n) \right|_{\zeta=0}.
\end{align}
Generators of a Lie group $G$ form a linear space called the Lie algebra.

In such a way generators $I_k$ are vectors collinear to the coordinate lines in a certain point. This feature allows one to use generators to define an exponential map on a group~\cite{JostJurgen1995Rgag,KuhnelWolfgang2006Dg:c}. The exponential map is a special coordinate frame on $G$ constructed as a projection the algebra onto the group. Element $g$ of the group $G$ with exponential (normal) coordinates $\theta$ is given by the following expression:
\begin{align}\label{exponential_map_definition}
    g(\theta_1, \cdots,\theta_n)=\exp\left[ \theta_k I_k \right]=1+ \theta_k I_k + \cfrac{1}{2!} (\theta_k I_k)^2 + \cdots. 
\end{align}
This expression means that an element of algebra $\theta_k I_k$ is projected on the group $G$, the right hand side of \eqref{exponential_map_definition} provides an explicit way to calculate $g$.

It should be highlighted that normal coordinates $\theta$ generated by $I_k$ which are defined via coordinate frame $\zeta$ do not match the coordinate frame $\zeta$. In other words, if generators $I_k$ are defined within frame $\zeta$ by 
\begin{align}
  I_k \overset{\text{def}}{=} \cfrac{\pd}{\pd \zeta_k} \left. g(\zeta) \right|_{\zeta=0},
\end{align}
then the correspondent normal coordinates $\theta$ do not match the original frame:
\begin{align}
  g(\zeta) \not = \exp \left[ \zeta_k I_k \right].
\end{align}
This is due to the fact that $I_k$ are first derivatives of $\zeta$, so the normal coordinates $\theta$ match coordinates $\zeta$ only in the linear order. Because of this in the most general case the normal coordinates may not cover the whole group $G$, but only a small neighbourhood around the identical transformations.

To break group $G$ spontaneously down to $H$ the following actions should be performed. First, the algebra of $H$ should be separated within algebra of $G$. Second, as generators of $H$ are vectors on the manifold $G$, they should be used to split $G$ into a series of layers. Namely, generators of $H$ should be treated as vectors tangent to a series of surfaces. This splits group $G$ in a series of layers with $H$ being a typical layer. Finally, generators of $G$ that do not belong to $H$ define directions orthogonal to these layers. Consequently, one can introduce a special coordinate frame that respect such a structure on $G$.

On the practical ground that coordinate frame is given by the following. We denote generators of $H$ as $V_k$; generators of $G$ which are orthogonal to $V_k$ we denote as $A_k$. An arbitrary element $g$ in such a coordinate frame is given by the following expression:
\begin{align}\label{group_parametrisation}
  g(\zeta, \phi) \overset{\text{def}}{=} \exp\left[ i \zeta_k A_k \right] \exp\left[ i \phi_k V_k \right].
\end{align}
Functions $\zeta$ and $\phi$ set a coordinate frame on $G$; functions $\phi$ define coordinate on layers, while coordinates $\zeta$ parameterise directions orthogonal to these layers. This allows one to study a structure of a smooth manifold $G/H$ which is covered with coordinates $\zeta$ and called a quotient of $G$ with respect to $H$.

An arbitrary element $C$ of the quotient $G/H$ reads
\begin{align}
  C(\zeta) \overset{\text{def}}{=} \exp\left[i \zeta_k A_k\right].
\end{align}
It should be noted that in such a definition all elements of $G/H$ belong to $G$, because $\zeta_k A_k$ still belongs to the algebra of $G$. A nonlinear action of group $G$ on its quotient $G/H$ is given by the following formula:
\begin{align}
  g ~\exp\left[i\zeta_k A_k\right] = \exp\left[ i \zeta'_k(\zeta,g) A_k \right]~\exp\left[i \phi_k(\zeta,g) V_k \right]. \label{action_on_a_quotient}
\end{align}

The geometrical meaning of formula~\eqref{action_on_a_quotient} should be understood as follows. On the left hand side of~\eqref{action_on_a_quotient} we have a multiplication of two group elements. This multiplication defines the action of an arbitrary element $g$ on an element $C$ from $G/H$. The result of this multiplication is another element of $G$ given in terms of normal coordinates~\eqref{group_parametrisation}. In other words, one shifts element $C$ via left multiplication on $g$ to a new position on the group manifold and evaluate its new coordinates.

Formula~\eqref{action_on_a_quotient} allows one to define two nonlinear realisations of $G$. First one is a direct action of $G$ on $G/H$ which is given in terms of $\zeta$ and $\zeta'$:
\begin{align}
    \zeta_k \to \zeta'_k(\zeta,g),
\end{align}
where $\zeta$ and $\zeta'$ are given by~\eqref{action_on_a_quotient}. For the purpose of our study we treat $\zeta$ as $\dim(G)-\dim(H)$ physical fields\linebreak $\zeta(t,x,y,z)$ that are transformed under a direct nonlinear action of $G$. The second realisation is a nonlinear action of $G$ on its subgroup $H$. It is defined on physical fields $\psi(t,x,y,z)$ that are already subjected to the group $H$:
\begin{align}\label{action_of_a_subgroup}
\psi \to \exp\left[ i u_k \hat{V}_k \right] \psi.
\end{align}
Here $\hat{V}_k$ are generators of $H$ in a suitable representation and $u_k$ are transformation parameters. Group $G$ acts on physical fields $\psi$ in a nonlinear way through the linear action of $H$ as follows:
\begin{align}
    \psi(t,x,y,z) \to \exp\left[ i\phi_k(\zeta(t,x,y,z),g) \hat{V}_k \right] \psi(t,x,y,z).
\end{align}
In this expression transformation parameters $\phi$ are given by formula~\eqref{action_on_a_quotient}, so they depend both on the bigger group element $g$ and physical fields $\zeta$. It should be noted that in order to construct such a nonlinear realisation of $G$ on $H$ it is required to have a nonlinear realisation of $G$ on $G/H$. At the same time a nonlinear realisation of $G$ on $G/H$ can be constructed without specifying a particular action of $G$ on $H$. Therefore we consider a nonlinear action of $G$ on $H$ as an auxiliary construction which may be included in a model.

The last thing required for a proper symmetry realisation are covariant derivatives. Introduction of covariant derivatives is necessary because of the nonlinear nature of transformations. In the most general case the standard kinetic term $\eta^{\mu\nu} \pd_\mu \zeta_k \pd_\nu \zeta_k$ is not invariant with respect to $G$ and cannot be used. Expressions for the covariant derivatives are given by the Cartan forms~\cite{Coleman:1969sm,Callan:1969sn,Volkov:1973vd}:
\begin{align}\label{covariant_derivatives}
    &
    \begin{cases}
    \nabla_\mu \zeta_k &= (P_k)_\mu , \\
    \nabla_\mu \psi &= \pd_\mu \psi + i (\Gamma_k)_\mu \hat{V}_k \psi, 
    \end{cases}
    \\
    & \omega_\mu \overset{\text{def}}{=} e^{-i\zeta_k A_k} \pd_\mu e^{i\zeta_k A_k}\overset{\text{def}}{=} i (P_k)_\mu A_k + i (\Gamma_k)_\mu V_k.
\end{align}
Formula \eqref{covariant_derivatives} provides the explicit way to obtain covariant derivatives which should be used to construct invariant Lagrangians.

The following feature of this technique should be highlighted. The nonlinear realisation is given by the corresponding transformations \eqref{action_on_a_quotient}, but the explicit form of the transformation is not required to construct an invariant Lagrangian. Formulae \eqref{covariant_derivatives} allow one to define kinetic terms for fields $\zeta_k$ and $\psi$ and to obtain an explicit form of an invariant Lagrangian. This feature is due as the structure of the derivatives respect the nonlinear symmetry realisation thereby it implicitly contains information on the transformations. For the sake of briefness we limit ourselves only to a discussion of covariant derivatives.

There are a few direct corollaries of this nonlinear realisation technique which we would like to highlight. The first one is the fact that the Goldstone theorem holds for such models (in full agreement with the logic presented in the previous section). Fields $\zeta_k$ are transformed in a nonlinear way, so in the most general case there is no ground to believe that $\zeta_k^2$ is invariant with respect to $G$. Therefore, in the most general case mass terms $\zeta_k^2$ must be excluded and fields $\zeta_k$ can be only massless. At the same time, fields $\zeta$ can only be bosons, which is also due to the nonlinear structure of the group action $G$. If $\zeta_k$ are fermion fields, then the standard kinetic term must be invariant with respect to $G$:
\begin{align}
\bar{\zeta}_k \gamma^\mu\nabla_\mu \zeta_k.
\end{align}
However $\zeta_k$ and $\nabla_\mu \zeta_k$ are transformed differently with respect to $G$, so such a term cannot be invariant\footnote{One may argue, that the term can be made invariant via a special transformation law for $\gamma^\mu$. In that case $\gamma$-matrices must have two spinor indices, that transforms differently with respect to $G$.}. These two statements form the Goldstone theorem and fields $\zeta_k$ are true massless goldstone bosons.

The other physical corollary is related to the structure of the quotient $G/H$ algebra. If algebra of $G/H$ is closed (i.e. all commutators of any two elements from the algebra lie in the algebra), then covariant derivatives of $\psi$ are trivial, i.e. they match the regular derivatives. This is due to the fact that the Cartan form is expressed in terms of commutators as follows:
\begin{align}
    e^{-i\zeta_k A_k} \pd_\mu e^{i\zeta_k A_k}=\sum\limits_{n=0}^\infty \cfrac{(-1)^n}{(n+1)!} \operatorname{ad}_{i\zeta_k A_k}^n ~ \{i ~ (\pd_\mu \zeta_k)~ A_k\}.
\end{align}
Here $\operatorname{ad}_X Y \overset{\text{def}}{=} [X,Y]$ is the adjoint action of the correspondent algebra. Therefore fields $\psi$ obtain a nontrivial covariant derivative if and only if $G/H$ algebra is not closed.

Finally, we would like to note that such a nonlinear realisation is related to a gauge field theory. We defined $\zeta_k$ as the propagating degrees of freedom and express forms $P$ and $\Gamma$ in terms of these fields. If we treat $(\Gamma_k)_\mu$ as the propagating degrees of freedom, then we no longer can use forms $(P_k)_\mu$ to define invariant quantities. At the same time we can define the standard Yang-Mills field tensor $F_{\mu\nu}\overset{\text{def}}{=}\pd_\mu \Gamma_\nu - \pd_\nu \Gamma_\mu + [\Gamma_\mu, \Gamma_\nu]$ which can be used to define the standard Yang-Mills kinetic term $\operatorname{tr}\left\{F_{\mu\nu} F^{\mu\nu}\right\}$ which is also invariant with respect to $G$. In such a way one can treat a nonlinear symmetry realisation as a generalisation of the standard gauge field theory. This generalisation replaces gauge vector bosons $\Gamma$ with a more fundamental degrees of freedom $\zeta$ which carry an additional symmetry.

In this section we presented one particular framework of nonlinear symmetry realisation. We implement this framework in the next sections and discuss physical features of gravity models with nonlinear realised symmetry.

\section{Nonlinear symmetry realisation for gravity}\label{Implementation}

As it was highlighted in the introduction, there are two ways to implement nonlinear symmetry realisation in a gravity model. First one is to implement it at the level of coordinates, second one is to implement it at the level of physical fields. Classical results \cite{Ogievetsky:1973ik,Borisov:1974bn} completely cover an opportunity of nonlinear symmetry realisation at the level of coordinates. Therefore this brunch of study lies beyond the scope of this article and further we address only an opportunity to introduce new symmetry at the level of physical fields.

To implement nonlinear symmetry realisation for gravity it is required to define a bigger group $G$ and a smaller group $H$. The logic presented in the previous section allows one to draw a few significant conclusions on physical properties of gravity models obtained in such a way.

First, in the previous sections it was proven that the Goldstone theorem holds for the nonlinear symmetry realisation, so the model must contain a certain number of massless goldstone bosons. In accordance with the empirical data~\cite{Abbott:2017vtc} we treat (low energy effective) gravitons as these goldstone modes. 

Second, gravitons are described within GR by the metric tensor $g_{\mu\nu}$ which has ten independent components. In order to have the same number of components it is required to use ten-dimensional Poincare group $\mathcal{P}$ as the small group $H$. Another reason to consider the Poincare group is the aforementioned Casimir operators, which allows one to consistently define mass and spin operators. Thus the choice of the smaller group $H$ is fixed uniquely.

Third, the choice of the bigger group $G$ is also heavily constrained. The simplest choice would be to take the bigger group $G$ as a direct product of the Poincare group $\mathcal{P}$ and an arbitrary Lie group such as $SU(N)$, but such a model cannot be considered satisfactory. In full agreement with the logic presented in the previous section the quotient $(\mathcal{P}\times SU(N))/\mathcal{P}$ is the $SU(N)$ group. Because of this the quotient $G/H$ has a closed algebra, so matter fields $\psi$ which are subjected to a nonlinear action of $G$ on $H$ obtain trivial covariant derivatives. To put it otherwise, in such a model goldstone bosons $\zeta_k$ that should be associated with the gravitational degrees of freedom do not interact with matter degrees of freedom that are subjected to $SU(N)$. This is the reason why such a choice cannot be considered interesting.

Finally, it is crucial to address the well-known theorem considering infinite-dimensional realisations of the Lorentz group. It is well-known that only infinite-dimensional representations of Lorentz group are unitary \cite{Weinberg:1995mt}. It may appear that this result constraints all possible nonlinear realisations at the level of field, but this is not the case. As it is pointed, for instance, in \cite{Weinberg:1995mt}, any theory must be unitary with respect to coordinate transformations. Therefore coordinate transformations must be subjected to an infinite-dimensional realisation of the Lorentz group. However, author of \cite{Weinberg:1995mt} also points out that ``There is no problem in working with non-unitary representations, because the objects we are now concerned with are fields, not wave functions, and do not need to have a Lorentz-invariant positive norm.''. Therefore one is free to consider any finite-dimensional nonlinear realisation of the Lorentz (or any bigger) group.

A proper way to define the bigger group $G$ is given by the Ogievetsky theorem~\cite{Ogievetsky:1973ik}. The theorem states that any generator of coordinate frame transformations can be obtained via commutation of generators from algebras of $C(1,3)$ and $SL(4,\mathbb{R})$. Therefore it is natural to consider either $C(1,3)$ or $SL(4,\mathbb{R})$ as perspective candidates for the bigger group $G$. We present algebras of $C(1,3)$ and $SL(4,\mathbb{R})$ in Appendix~\ref{appendix_algebras}. 

Results of \cite{Borisov:1974bn} serve as additional justification of usage of $C(1,3)$ and $SL(4,\mathbb{R})$ algebras. In paper \cite{Borisov:1974bn} it was shown that the whole Riemannian geometry is reproduced via a combined nonlinear realisation of $C(1,3)$ and $SL(4,\mathbb{R})$ algebras at the level of coordinates. It is natural to expect a nonlinear realisation of the same symmetries at the level of physical fields. Therefore, paper \cite{Borisov:1974bn} provides a result fundamental for all further studies of nonlinear symmetry realisation in gravity. The cornerstone of this result is the usage of two groups which algebras cannot be merged in a closed algebra. The goal we pursuit is to establish whether models with the standard nonlinear realisation (at the level of physical fields) of these closed algebras are relevant for gravity studies. Further in this paper we constrain ourselves only to models with closed algebras.

It should also be noted that similar attempts to construct a model with a nonlinearly realised conformal symmetry took place before \cite{Isham:1970gz,Isham:1971dv,Isham:1970xz}. Namely, paper \cite{Isham:1971dv} partially reproduces results presented in \cite{Borisov:1974bn}. Model presented in \cite{Isham:1970gz} uses a different nonlinear realisation technique. Finally, paper \cite{Isham:1970xz} address a combined realisation of conformal and chiral symmetries. These results can hardly be considered relevant for our study, as they either deal with non-minimal models, or address issues lying beyond the scope of this paper.

Accounting for this line of argumentation we proceed with a construction of a nonlinear symmetry realisation.
Algebras of $C(1,3)$ and $SL(4,\mathbb{R})$ can extend the\linebreak Poincare algebra with three families of operators: $R_{(\mu)(\nu)}$, $K_{(\mu)}$, and $D$. However, some of these extensions are irrelevant for our study. It should be highlighted that we put generators indices in brackets to distinguish them from the Lorentz indices. It is impossible to include both $R_{(\mu)(\nu)}$ and $K_{(\mu)}$ in an extended algebra since, as it is pointed in~\cite{Ogievetsky:1973ik}, an algebra that contains both of these operators is not closed. Despite the fact that it is possible to study a nonlinear realisation of such an algebra~\cite{Borisov:1974bn}, such a study lies beyond the area of applicability of the method presented in the previous section. Moreover, if we only add either operators $K_{(\mu)}$ or operator $D$, then we create a model with trivial covariant derivatives. This is due to the fact that algebras of $D$ and $K_{(\mu)}$ are closed and, in full analogy with the previous case, goldstone bosons that should be associated with the gravitational degrees of freedom are decoupled from the regular matter.

This logic allows us to find three extension of the Poincare group that could not be immediately dismissed. These extensions are:
\begin{enumerate}
\item
    extension of the Poincare algebra with operators $D$ and $K_{(\mu)}$,
\item
    extension of the Poincare algebra with operators $R_{(\mu)(\nu)}$,
\item
    extension of the Poincare algebra with operators $R_{(\mu)(\nu)}$ and $D$.
\end{enumerate}

Another important feature of models with such a nonlinear realisation of symmetry is related to a special role of scalar fields. Covariant derivatives of matter fields (which are subjected to the linear Poincare group action) are given by the general formula~\eqref{covariant_derivatives}:
\begin{align}
\nabla_\mu \psi = \pd_\mu\psi + i (\Gamma_k)_\mu \hat{V}_k \psi.
\end{align} 
In this formula $\hat{V}_k$ corresponds to the Poincare group generators. In the case of scalar fields, the correspondent Poincare generators vanish and scalar fields always have trivial covariant derivatives. At the same time, it is possible to introduce scalar fields that are subjected to a nonlinear action of $G$ on $G/H$. These fields can obtain a nontrivial covariant derivative. Therefore such models do have a room for scalar fields interacting with gravity, but the role of scalar fields is significantly different from the standard treatment.

The simplest expansion of the Poincare algebra is given by the extension with operators $D$ and $K_{(\mu)}$. 
This model serves as, perhaps, the simplest modification of GR induced via nonlinear symmetry realisation. The correspondent quotient algebra is closed and the matter fields have trivial covariant derivatives.
Because of this the Lorentz group is realised in a linear way that matches GR exactly and no modification to the Einstein sector of gravity can be made.
Nonetheless, nonlinear realisation introduces five additional scalar degrees of freedom with non-trivial interactions. Wide phenomenology of scalar-tensor models allows one to conjecture that their existence can result in a natural appearance of inflation \cite{Clifton:2011jh}. A more detailed study of this model lies beyond the scope of this paper and will be discussed elsewhere.
We discuss this model for the sake of illustration, as it has a few properties common to all models with nonlinear symmetry realisation. New operators $D$ and $K_{(\mu)}$ generate five new physical fields $\phi$ and $\sigma^{(\alpha)}$. It should be highlighted, that the index $(\alpha)$ is not a Lorentz index, so fields $\sigma^{(\alpha)}$ are four scalar fields subjected to nonlinear transformations.

The correspondent covariant derivatives read:
\begin{align}
    \begin{cases}
    \nabla_\mu \phi =\pd_\mu \phi, \\
    \nabla_\mu \sigma^{(\alpha)} =\cfrac{e^\phi-1}{\phi}~ \pd_\mu \sigma^{(\alpha)} -\cfrac{e^\phi-\phi-1}{\phi^2}~ \pd_\mu\phi~ \sigma^{(\alpha)}.
    \end{cases}
\end{align}
Nonlinear transformations of these fields are given by the following expression:
\begin{align}
\exp\left[\cfrac{i}{2}\,\theta^{(\mu)(\nu)} \, L_{(\mu)(\nu)} + i \theta D + i \theta^{(\mu)} K_{(\mu)}\right] \exp\left[i\,\phi D + i\, \sigma^{(a)} K_{(a)}\right] = \nonumber \\
=\exp\left[i\,\phi' D + i\,\sigma'{}^{(a)} K_{(a)}\right] \exp\left[ \cfrac{i}{2}\,u^{(\mu)(\nu)} L_{(\mu)(\nu)} \right] ~.
\end{align}
Here $\theta^{(\mu)(nu)}$, $\theta^{(\mu)}$, and $\theta$ are the transformation parameters; $\phi$, $\sigma^{(a)}$ and $\phi'$, $\sigma'{}^{(a)}$ are fields before and after the transformations; $u^{(\mu)(\nu)}$ are parameters defining the nonlinear group action on the standard field subjected to Lorentz transformations.

In these expressions fields $\sigma^{(\alpha)}$ can have a canonical mass dimension, while field $\phi$ should be dimensionless. In order to use variables with the canonical mass dimension we introduce an energy scale $\ce$ and define a field $\psi$ with the canonical mass dimension:
\begin{align}
\psi=\ce \phi.
\end{align}
The energy scale $\ce$ should be treated as a symmetry breaking scale. Such a construction allows one to write the Lagrangian of the system:
\begin{align}
    \mathcal{L} =& \cfrac12 (\nabla\psi)^2 +\cfrac12 (\nabla\sigma)^2 =\label{Largangian_1} \\
    =& \cfrac12 \left[ 1+\cfrac{\sigma^2}{\ce^2} \left( \cfrac{e^{\psi/\ce}-\psi/\ce-1}{(\psi/\ce)^2} \right)^2 \right] (\pd\psi)^2 +\cfrac12 \left( \cfrac{e^{\psi/\ce}-1}{\psi/\ce} \right)^2 (\pd \sigma)^2 \nonumber \\
& \hspace{3cm}-\cfrac{1}{\ce} \cfrac{e^{\psi/\ce}-1}{\psi/\ce}\cfrac{e^{\psi/\ce}-\psi/\ce-1}{(\psi/\ce)^2} \pd_\mu\psi \pd^\mu \sigma^{(\alpha)} \sigma_{(\alpha)} = \nonumber \\
  =&\cfrac12 (\pd\psi)^2 + \cfrac12 (\pd \sigma)^2 + \cfrac{1}{2\ce} \left( \psi (\pd 
  \sigma)^2 - \pd_\mu\psi \pd^\mu \sigma^{(\alpha)} \sigma_{(\alpha)} \right)\nonumber \\
& \hspace{3cm} +\cfrac{1}{\ce^2} \left[ \cfrac18 (\pd \psi)^2 \sigma^2 + \cfrac{7}{24} \psi^2 (\pd \sigma)^2 -\cfrac{5}{12} \phi \pd_\mu\phi \pd^\mu \sigma^{(\alpha)} \sigma_{(\alpha)} \right] + \cdots .\nonumber
\end{align}

This Lagrangian contains an infinite number of interaction terms, however, higher interaction terms are suppressed by higher powers of the symmetry breaking scale. As we about to see, such a behaviour is typical for models of that kind. This feature of models under consideration goes in line with the assumption that the Planck mass should be treated as the proper symmetry breaking scale.

It should be noted that even such a simple model can have non-trivial implications for gravity theory. Namely, the amount of new degrees of freedom introduced in this model is sufficient to define the metric-independent volume element constructed in \cite{Guendelman:1999qt}. Consequently, even within such a simple model it may be possible to associate the goldstone degrees of freedom appearing due to the nonlinear symmetry realisation with the physical volume and, therefore, with the spacetime geometry. A more detailed discussion of this opportunity lies beyond the scope of this paper and are to be pursued elsewhere.

A more sophisticated model is based on the extension of the Poincare algebra with operators $R_{(\mu)(\nu)}$.
In this paper we follow the standard approach \cite{Weinberg:1978kz,Coleman:1969sm,Callan:1969sn} and associate coset coordinated $h^{(\mu)(\nu)}$ conjugated with $R_{(\mu)(\nu)}$ with small metric perturbations. It should be highlighted, that the classical paper \cite{Borisov:1974bn} follows a different approach. As we highlighted before, the approach of \cite{Borisov:1974bn} pursuits the goal to restore GR, so it is justified for this particular problem, but it should not be used in the context of our paper, as we pursuit to study an alternative approach to the same problem lying beyond the one taken in \cite{Borisov:1974bn}.
The structure of the Cartan form provides a simple way to study the structure of interaction in this model:
\begin{align}
& \exp\left[-\frac{i}{2} h^{(\mu)(\nu)} R_{(\mu)(\nu)}\right] d \exp\left[\frac{i}{2} h^{(\mu)(\nu)} R_{(\mu)(\nu)}\right] = \\
=& R_{(\mu)(\nu)} \left\{ \cfrac{i}{2} d h^{(\mu)(\nu)} + \sum\limits_{n=1}^\infty \cfrac{1}{(2n+1)!} \left(\cfrac{-i}{2}\right)^{2n} \left(\ad^{2n}_h dh \right)^{(\mu)(\nu)} \right\}\nonumber \\
& - L_{(\mu)(\nu)} \left\{ \sum\limits_{n=1}^\infty  \cfrac{1}{(2n)!} \left(\cfrac{-i}{2}\right)^{2n-1} \left( \ad^{2n-1}_h dh \right)^{(\mu)(\nu)} \right\} \nonumber = \\
=&\cfrac{i}{2} \left[ dh^{(\alpha)(\beta)} - \cfrac13 h^{(\alpha)(\lambda)} h^{(\beta)(\sigma)} dh^{(\sigma)(\lambda)}  + \cfrac13 h^{(\alpha)(\sigma)} h^{(\sigma)(\lambda)} dh^{(\lambda)(\beta)} + O(h^5) \right] R_{(\alpha)(\beta)} \nonumber \\
& + \cfrac{i}{2} \left[ h^{(\alpha)(\lambda)} dh^{(\lambda)(\beta)} + O(h^4) \right] L_{(\alpha)(\beta)} . \nonumber
\end{align}

The first important feature of this model is the fact that matter fields do obtain nontrivial covariant derivatives and do interact with the gravitational degrees of freedom. However, this interaction has a specific form which can be seen from the correspondent part of the Cartan form:
\begin{align}
\left\{ \sum\limits_{n=1}^\infty  \cfrac{1}{(2n)!} \left(\cfrac{-i}{2}\right)^{2n-1} \left( \ad^{2n-1}_h dh \right)^{(\mu)(\nu)} \right\} =  \cfrac{i}{2} \left[ h^{(\alpha)(\lambda)} dh^{(\lambda)(\beta)} + O(h^4) \right].
\end{align}
This expression defines the structure of the matter field covariant derivatives and the leading interaction term is quadratic in $h$. Moreover, this expression contains only even powers of $h$. Therefore, within this model there is only room for interaction of an even number of fields $h$ with the matter fields.

To understand this feature it is required to analyse self-interaction of $h$. It can be seen that fields $h$ are dimensionless, so in full analogy with the previous case we introduce a mass parameter $m$ that plays a role of the symmetry breaking scale and define a field variable with the canonical mass dimension
\begin{align}
k^{(\mu)(\nu)} = m ~ h^{(\mu)(\nu)}.
\end{align}
The Lagrangian of the goldstone modes $h$ reads
\begin{align}
    &\mathcal{L}_h =\cfrac12 ~ \nabla_\mu h^{(\alpha)(\beta)} \nabla^\mu h^{(\alpha)(\beta)}= \\
    &= \cfrac12~ \pd_\mu k^{(\alpha)(\beta)} ~ \pd^\mu k^{(\alpha)(\beta)} -\cfrac{1}{6 m^2} \left[~\pd_\mu k^{(\alpha)(\beta)} \pd^\mu k^{(\rho)(\sigma)} ~ k^{(\rho)(\alpha)} k^{(\sigma)(\beta)} - \pd_\mu k^{(\alpha)(\sigma)} k^{(\sigma)(\lambda)} k^{(\lambda)(\rho)} \pd^\mu k^{(\rho)(\alpha)} \right] \nonumber\\
& \hspace{13cm}+ O(m^{-4}). \nonumber
\end{align}
As it was highlighted before, the model has an infinite number of self-interaction terms, but higher interaction terms are suppressed by higher powers of the symmetry breaking scale. The other thing that should be noted is the fact that such a Lagrangian also contains only even powers of fields $h$.

These results should be understood as follows. We define fields $h$ as degrees of freedom carrying nonlinear symmetry realisation of the bigger group. The correspondent invariant Lagrangian of these degrees of freedom is subjected to the Goldstone theorem and provides a consistent description of self-interaction of these fields. Introduction of the new symmetry resulted in the fact that only even number of fields $h$ can interact either with each other or with the regular matter.

This feature of the model is exotic since within GR gravity manifests other features. First, as we highlighted before, all matter scalar fields do not interact with gravity within this model. 
This feature is due to the structure of the covariant derivative \eqref{covariant_derivatives}. The interaction with field $h_{\mu\nu}$ is contained in form $(\Gamma_k)_\mu$. This form is coupled to an arbitrary field via a Lorentz group generator $\hat{V}_k$. In case of a scalar field the generator vanishes identically, thus the covariant derivative matches the standard derivative. Consequently, any scalar field cannot be coupled to tensor field $h_{\mu\nu}$ via the covariant derivative. Nonetheless, a scalar field can be coupled to $h_{\mu\nu}$ in the same way it is coupled to the standard small metric perturbations within GR. Because of this a further more detailed investigation of interaction between $h_{\mu\nu}$ and scalar field is required.
Second, standard gauge vector (massless) fields interact with gravity in a manner similar to GR. Namely, the standard gauge field kinetic term $g^{\mu\nu} g^{\alpha\beta} \operatorname{tr} \{F_{\mu\nu} F_{\alpha\beta}\}$ is quadratic in gravitational interaction in the lowest order, although it contains interaction terms with an odd number of gravitons. Finally, within GR gravitational interaction between fermions can only be given in terms of the vierbein (tetrad) formalism. The method of nonlinear symmetry realisation presented in this paper has no obvious analogy of the vierbein formalism. Therefore gravitational interaction between fermions within this model requires more fine tools of analysis.

Such a behaviour of the goldstone modes may be interpreted as follows. One should establish a distinction between states with a well-defined symmetry and well-defined interactions. The standard GR gravitons that admin linear interaction with matter should be considered as states with well-defined interaction properties. States described by fields $h$ should be considered as states with well-defined symmetry properties, but lacking a simple form of interaction. Such a treatment is similar to the one used for neutrino mixing. In a similar way one defines neutrino states with well-defined mass $(\nu_1,\nu_2,\nu_3)$ which are states with well defined Poincare symmetry and complement them with states with well-defined interaction $(\nu_e,\nu_\mu,\nu_\tau)$ which are states with the well-defined $SU(2)$ symmetry.

The last model that is relevant to us is an extension of the Poincare algebra with operators $R_{(\mu)(\nu)}$ and $D$. This model is reduced to the previous one almost completely. The reason behind this is the fact that the operator $D$ commutes with $R_{(\mu)(\nu)}$ and $L_{(\mu)(\nu)}$. Because of this the Cartan form is simplified:\hfill
\begin{align}
\omega_\mu &= \exp\left[-\frac{i}{2} h^{(\mu)(\nu)} R_{(\mu)(\nu)} -i \phi D\right] \pd_\mu \exp\left[\frac{i}{2} h^{(\mu)(\nu)} R_{(\mu)(\nu)} +i \phi D\right] =\\
&= i ~ d\phi ~D + \exp\left[-\frac{i}{2} h^{(\mu)(\nu)} R_{(\mu)(\nu)}\right] \pd_\mu \exp\left[\frac{i}{2} h^{(\mu)(\nu)} R_{(\mu)(\nu)}\right]. \nonumber
\end{align}
Therefore the introduction of $D$ together with $R_{(\mu)(\nu)}$ simply introduces an additional scalar field that is completely decoupled from gravitational degrees of freedom.

Finally, it should be highlighted, that there is no obvious way to show that variable $h_{\mu\nu}$ contains only spin $2$ components. In the most general case a tensor field $h_{\mu\nu}$ can contain spin $1$ and spin $0$ components alongside spin $2$ modes. These modes are to affect gravitational phenomena at all spatial and energy scales. Consequently, existence (or non-existence) of auxiliary gravity modes in the model serves as an important test of the model. Within GR the auxiliary modes are excluded due to the gauge symmetry. In the proposed model the symmetry is more sophisticated, therefore there is no simple way to exclude the new modes. This feature of the model is to be studied elsewhere.

This array of results is summarised in the next section together with conclusions.

\section{Summary and Conclusion}\label{Discussion}

The results presented in the previous sections can be summarised as follows.

The known classical results were discussed in Sections~\ref{Introduction} and~\ref{many_faces_of_nonlinear_symmetry_realisation} in order to setup the problem and the approaches to solve it. The search for modified models of gravity is mainly performed via modification of GR Lagrangian. Despite the fruitfulness of that approach it is not suitable for studying the role of models symmetries. We pursue the idea to study opportunities to create modified gravity models via introduction of a new symmetry to the gravitational sector.

The opportunity to introduce a new symmetry is constrained by the fact that gravitational field defines properties of the spacetime. Because of this the only valuable option is a nonlinear realisation of an additional symmetry. 
Namely, we pursuit nonlinear realisation of additional symmetry at the level of physical field, as we discuss in Section \ref{Introduction}.
As we discussed in the Section~\ref{many_faces_of_nonlinear_symmetry_realisation} nonlinear symmetry realisation is the same phenomenon as spontaneous symmetry breaking. Therefore, the Goldstone theorem holds and massless goldstone bosons can be identified with gravitons.

The idea of spontaneous symmetry breaking for gravity is natural in the following context. There is a number of theoretical obstacles preventing one from a construction of a quantum theory of gravity. Despite this fact a consistent quantum treatment of gravity in the low energy regime is possible within the effective field theory formalism. This formalism describes all observed phenomena in terms of low energy effective gravitons and their interactions. The consistent treatment of GR as an effective theory provides a ground to conjecture that the low energy gravity behaviour is due to a spontaneous breaking of some symmetry. Therefore low energy effective gravitons which are massless goldstone bosons can appear due to the Goldstone theorem. GR has a single mass parameter, namely, the Planck mass which in this case should be considered as the symmetry breaking scale (or directly related to it). Therefore it is reasonable to search for an opportunity to implement nonlinear realisation of some symmetry in the gravitational sector.

The next important issue discussed in this paper is related to the opportunity to create a valuable model of gravity with a spontaneously broken symmetry and it is presented in Sections~\ref{Nonlinear_symmetry_realisation_quotient} and~\ref{Implementation}. 
It is possible to define states with well-defined mass and spin in the observable universe, which is due to the Poincare symmetry of the observed spacetime. And it is possible to preserve that feature, if a gravity model experiences spontaneous symmetry breaking down to the Poincare group. Therefore it is only possible to choose a bigger symmetry group which is broken spontaneously. The choice of the bigger group is also constrained, as models with nonlinear symmetry realisation share a few important features.

Finally, the role of classical results \cite{Borisov:1974bn} is discussed in section \ref{Implementation}. The classical result is based on a combined realisations of groups $C(1,3)$ and $SL(4,\mathbb{R})$. It provides one with framework of Riemannian geometry restored via nonlinear symmetry realisation technique. This result is central for nonlinear symmetry realisation studies and its role should not be overlooked. In the context of our study it proves two important features of nonlinear symmetry realisation. 
First, nonlinear symmetry realisation does take place at the level of coordinate transformations. 
Second, it is required to use a combined nonlinear realisation of groups $C(1,3)$ and $SL(4,\mathbb{R})$ to restore Riemannian geometry. 
This provides one with a ground to expect existence of a nonlinear realisation of the same symmetries at the level of physical fields.
Combination of $C(1,3)$ and $SL(4,\mathbb{R})$ algebras doesn't form a closed algebra, so the standard nonlinear symmetry realisation technique (implemented at the level of physical fields) cannot be used. Because of this we use the standard nonlinear realisation technique which is, in some sense, complimentary to the one presented in the classical paper. Namely, we only consider extensions of Poincare algebra that are closed, so the standard nonlinear symmetry realisation formalism can be applied. Therefore we do not expect to restore general relativity, but to obtain alternative gravity models that may be relevant for further studies.

An important result presented in this paper is the fact that gravity models with the particular implementation of nonlinear symmetry breaking cannot describe interaction of the standard scalar particles (the one subjected to a linear action of the Poincare group) with the goldstone modes. At the same time scalar particles subjected to a nonlinear action of the bigger group may participate in gravitational interactions.

We obtained two minimal gravity models based on principles presented above. The first model is given by Lagrangian \eqref{Largangian_1} and belongs to scalar-tensor models of gravity. Models of these type have a wide phenomenology. For instance, the new scalar degrees of freedom may drive inflation \cite{Clifton:2011jh} thereby describing it as an effect induced by spontaneous symmetry breaking.

The second model is based on a nonlinear realisation of the Poincare algebra extended with operators $R_{(\mu)(\nu)}$ from $SL(4,\mathbb{R})$. This model has a symmetric matrix $h^{(\mu)(\nu)}$ that describes massless goldstone bosons that appear due to the spontaneous symmetry breaking. The model contains an infinite number of interaction terms, but higher order interaction terms are suppressed by higher powers of the symmetry breaking scale (the Planck mass). At the same time due to the existence of a new symmetry the model only contains interaction of even number of the goldstone fields. As all models of that class the model cannot describe interaction between the goldstone modes and the scalar particles. Interaction between spinor and vector particles require at least two goldstone particles. Such an interaction is exotic since within GR fermions can interact with a single graviton.

We propose the following interpretation of that phenomenon. We treat states describe by $h^{(\mu)(\nu)}$ as states with a well-defined symmetry. The standard GR gravitons, i.e. states that interact with matter particles, should be treated as states with well-defined interactions. In such a way the proposed model may describe gravity in terms of states with a well-defined symmetry, which are related to states with well-defined interactions in a nonlinear way. Such a mechanism is analogous to the well-known mechanism of neutrino mixing. Gravitational states with a well-defined symmetry and well-defined interactions should be distinguished in the same way one distinguishes neutrino with a well-defined mass (well-defined Poincare symmetry) and well-defined interactions (well-defined $SU(2)$ symmetry).

In conclusion we would like to highlight that a further study of the proposed model is required. The fact, that gravitational states with a well-defined symmetry and well-defined interactions are connected in a non-linear way makes treatment of the model more complicated. Aforementioned relation between graviton states might complexify perturbative treatment of the model.


\appendix

\section{Algebras of $C(1,3)$ and $SL(4,\mathbb{R})$}\label{appendix_algebras}

Generators of $C(1,3)$ are defined as follows:
\begin{align}
  \begin{cases}
    L_{\mu\nu} = x_\mu P_\nu - x_\nu P_\mu , \\
    P_\mu = i \pd_\mu , \\
    K_\mu = 2 x_\mu D - x^2 P_\mu, \\
    D =x^\mu P_\mu .
  \end{cases}
\end{align}
They form the following algebra:\hfill
\begin{align}
\begin{cases}
  [P_\mu,P_\nu]=0, \\
  [K_\mu,K_\nu]=0, \\
  [P_\mu,D]=i P_\mu , \\
  [K_\mu,D]=-i K_\mu,  \\
  [P_\mu,K_\nu]=2i(\eta_{\mu\nu} D - L_{\mu\nu} ), \\
  [L_{\mu\nu},D]=0 , \\
  [L_{\mu\nu},K_\alpha]=-i (\eta_{\mu\alpha} K_\nu - \eta_{\nu\alpha} K_\mu ), \\
  [L_{\mu\nu},P_\alpha]=-i(\eta_{\mu\alpha} P_\nu -\eta_{\nu\alpha}P_\mu), \\
  [L_{\mu\nu},L_{\alpha\beta}] = -i \left( L_{\mu\alpha}\eta_{\nu\beta} -L_{\mu\beta}\eta_{\nu\alpha} +L_{\nu\beta}\eta_{\mu\alpha} - L_{\nu\alpha}\eta_{\mu\beta} \right). 
\end{cases}
\end{align}

Generators of $SL(4,\mathbb{R})$ are
\begin{align}
  \begin{cases}
    R_{\mu\nu} = x_\mu P_\nu + x_\nu P_\mu , \\
    L_{\mu\nu} = x_\mu P_\nu - x_\nu P_\mu .
  \end{cases}
\end{align}
They form the following algebra
\begin{align}
\begin{cases}
  [L_{\mu\nu}, L_{\alpha\beta}] &= -i (L_{\mu\alpha} \eta_{\nu\beta} + L_{\nu\beta} \eta_{\mu\alpha}-L_{\mu\beta}\eta_{\nu\alpha} - L_{\nu\alpha} \eta_{\mu\beta} ) , \\
  [R_{\mu\nu}, R_{\alpha\beta}]&=i (L_{\mu\alpha} \eta_{\nu\beta} + L_{\nu\beta} \eta_{\mu\alpha}+L_{\mu\beta}\eta_{\nu\alpha} + L_{\nu\alpha} \eta_{\mu\beta} ), \\
  [L_{\mu\nu},R_{\alpha\beta}] &= i (R_{\mu\alpha} \eta_{\nu\beta} - R_{\nu\beta} \eta_{\mu\alpha}+R_{\mu\beta}\eta_{\nu\alpha} - R_{\nu\alpha} \eta_{\mu\beta} )
\end{cases}
\end{align}

The composed algebra of $C(1,3)$ and $SL(4,\mathbb{R})$ is not closed and admits the following relations:
\begin{align}
    \begin{cases}
        [L_{\mu\nu},D] &=0, \\
        [R_{\mu\nu},D] &=0, \\
        [R_{\mu\nu},P_\alpha] &=-i(\delta_{\mu\alpha} P_\nu + \delta_{\nu\alpha} P_\mu),\\
        [R_{\mu\nu},K_\alpha] &=-4 i x_\mu x_\nu P_\alpha + i (2 x_\mu D + x^2 P_\mu) \delta_{\nu\alpha} + i (2x_\nu D + x^2 P_\nu) \delta_{\mu\alpha}.
    \end{cases}
\end{align}

\bibliographystyle{unsrturl}
\bibliography{Biblio}

\end{document}